\documentclass{article}
\usepackage{amsfonts}
\usepackage[utf8]{inputenc}
\usepackage{amsmath, amssymb, amsthm, natbib, bm}
\usepackage[basic]{complexity}
\usepackage{algorithm2e}
\usepackage{graphicx}
\usepackage{pgfplots}
\usepackage{tikz}
\usepackage{color}
\usepackage{makeidx}
\usepackage{xcolor}
\makeindex

\usetikzlibrary{shapes, patterns, decorations, fit, intersections}
\newtheorem{theorem}{Theorem}

\newtheorem{definition}[theorem]{Definition}

\newcommand{\idf}[2]{\textit{#1}\index{#2}}

\newcommand*{\dotprod}[2]{\langle#1,#2\rangle}

\newcommand*{\abs}[1]{\lvert#1\rvert}

\newtheorem{assumption}[theorem]{Assumption}

\def\uniset{{\rm 1\kern-.40em 1}}

\input{tcilatex}

\begin{document}

\title{{\bf Structural Estimation of Matching Markets with Transferable Utility}\footnote{This paper is to be published by Cambridge University Press in the volume \emph{Online and Matching-Based Market Design} edited by Federico Echenique,
Nicole Immorlica, and Vijay Vazirani (2022). This arxiv version is not for distribution or use in derivative works. We thank Nikhil Agarwal and Paulo Somaini for their comments and Gabriele Buontempo for  superb research assistance.}}
\author{Alfred Galichon\footnote{New York University and Sciences Po. Support from ERC grant EQUIPRICE No. 866274 is acknowledged.} \and Bernard Salani\'e\footnote{Columbia University.}}
\date{\today }
\maketitle

\label{ch:structuraltumodels}

In matching models with transferable utility, the partners in a match
agree to transact in exchange of a transfer of num\'eraire (utility or money) from one side of the match to the other. While transfers may be non zero-sum (if for instance there is diminishing marginal utility, frictions, or other costs) or constrained, we focus in this chapter on the simplest case of \idf{perfectly transferable utility}{perfectly transferable utility}, in which the transfers are   unlimited and  zero-sum: the transfer agreed to by one partner is fully appropriated by the other side.
For simplicity, we also limit our discussion to the one-to-one bipartite
model: each match consists of two partners, drawn from two separate
subpopulations. The paradigmatic example is the
\idf{heterosexual marriage
market}{Heterosexual marriage market}, in which the two subpopulations are
men and women. We will use these terms for concreteness.

 With perfectly transferable utility, the main object of interest is the \idf{joint surplus function}{Joint surplus function}. It maps the characteristics of a man and a woman into the surplus utility created by their match, relative to the sum of the utilities they would achieve by staying single.  Knowing the joint surplus function is informative about the preferences of the partners, and about their interaction within the match. It also opens the door to counterfactual analysis, for instance of the impact of policy changes.

We assume that the  analyst observes a discrete set of characteristics
for each individual: their education, their age, their income category, etc.
Each combination of the values of these characteristics defines a \emph{type}. In any real-world
application, men and women of a given observed type will also vary in their
preferences and more generally in their ability to create joint surplus in
any match. We will assume that all market participants observe this additional variation, so that it contributes in determining the observed matching.  On the other hand, by definition it constitutes \idf{unobserved heterogeneity}{Unobserved heterogeneity} for the analyst.  The main challenge in this field is to recover the parameters of the joint surplus function without restricting too much this two-sided unobserved heterogeneity.

Matching with transferable utility solves a linear programming problem. In
recent years it has been analyzed with the methods of \idf{optimal
transport}{Optimal transport}. Under an additional \textquotedblleft
\idf{separability}{Separability}\textquotedblright\ assumption, most functions of interest are
convex; then \idf{convex duality}{Convex duality} gives a simple and transparent path to
identification\idf{identification}{Identification} of the parameters of these models\footnote{%
We collected the elements of convex analysis used in this chapter in
Appendix~A.}. The empirical implementation is especially straightforward when
the unobserved heterogeneity has a  \idf{logit}{Logit} form and the joint
surplus is linear in the parameters. Then the parameters can be estimated by
minimizing a globally convex objective function.

Section~\ref{ch:structuraltumodels:sec:unobsh} introduces  separable matching models.  Section~\ref{ch:structuraltumodels:sec:ident} presents assumptions under which data on ``who matches whom'' (the \idf{matching patterns}{Matching patterns}) identifies the parameters of the joint surplus function, and possibly also of the distributions of unobserved heterogeneity. We will also show how these parameters can be estimated (Section~\ref{ch:structuraltumodels:sec:estimation}),  and how to compute the stable matchings for given parameter values (Section~\ref{ch:structuraltumodels:sec:computation}).

\medskip

\textbf{Notation}. We use bold letters for vectors and matrices.  For any doubly-indexed variable
$\bm{z}=(z_{ab})$, we use the notation $\bm{z}_{a\cdot }$ to denote the vector of values
of $z_{ab}$ when $b$ varies; and we use a similar notation for $\bm{z}_{\cdot b}$.

\section{Matching with unobserved heterogeneity}\label{ch:structuraltumodels:sec:unobsh}

\subsection{Population and preferences}
We consider a population of men indexed by $i$ and a population of women indexed by $j$. Each match must consist of one man and one woman; and individuals may remain single.
If a man $i$  and a woman $j$ match, the assumption of
\idf{perfectly transferable utility}{Perfectly transferable utility} implies that their respective utilities can
be written as
\begin{align*}
& \alpha _{ij}+t_{ij} \\
& \gamma _{ij}-t_{ij}
\end{align*}%
where $t_{ij}$ is the (possibly negative) transfer from $j$ to $i$\footnote{If $t_{ij}$ is negative, it should be interpreted as a transfer of $-t_{ij}$ from $i$ to $j$. Also, $t_{i0}=t_{0j}=0$}.   Transfers can take all values on the real line, and are costless.  We assume that each individual knows the equilibrium values
of the transfers for all matches that (s)he may take part in, as well as his/her pre-transfer utility $\alpha_{i\cdot}$ or $\gamma_{\cdot j}$.

One key feature of markets with perfectly transferable utility is that  matching patterns do not depend on $\bm{\alpha}$ and $\bm{\gamma}$ separately, but only on their sums, which we call the \idf{{\em joint surplus}}{Joint surplus}\footnote{Strictly speaking, it only  is  a ``surplus'' when all $\alpha_{i0}$ and $\gamma_{0j}$ are zero. We follow common usage here.}.
\begin{definition}[Joint Surplus]\label{ch:structuraltumodels:def:jointsurp}
The joint surplus of a match is the sum of (pre- or post-transfers) utilities:
\[
    \tilde{\Phi}_{ij} = (\alpha_{ij}+t_{ij})+(\gamma_{ij}-t_{ij})=\alpha_{ij}+\gamma_{ij}.
    \]
    We extend the definition to singles with $\tilde{\Phi}_{i0}=\alpha_{i0}$ and $\tilde{\Phi}_{0j}=\gamma_{0j}$.
\end{definition}
To see this, note  that any change
\[
    (\alpha_{ij}, \gamma_{ij}) \to (\alpha_{ij}-\delta, \gamma_{ij}+\delta)
    \]
can be neutralized by adding $\delta$ to the transfer $t_{ij}$. This combined change leaves post-transfer utilities unchanged; therefore it does not affect the decisions of the market participants.

A {\em matching\/} is simply a set $\bm{d}$ of 0--1 variables $(d_{ij})$ such that $d_{ij}=1$ if and only if $i$ and $j$ are matched, along with 0--1 variables $d_{i0}$ (resp.\ $d_{0j}$) that equal~1 if and only if man $i$ (resp.\ woman $j$) is unmatched (single). It is {\em feasible\/} if no partner is matched more than once:

for all $i$, $\sum_j d_{ij} + d_{i0}=1$; and for all $j$, $\sum_i d_{ij}+d_{0j}=1$.

\subsection{Stability}\label{ch:structuraltumodels:sub:stabil}
Our notion of equilibrium is \idf{stability}{Stability}. Its definition in the context of models with perfectly transferable utility  is as follows\footnote{It can be seen as a special case of the more general definition of stability.}.
\begin{definition}[Stability---primal definition]\label{ch:structuraltumodels:def:stability:ptu1}
A feasible matching  is stable if and only if
\begin{itemize}
    \item no match has a partner who would rather be single
    \item no pair of currently unmatched partners would rather be matched.
\end{itemize}
    \end{definition}
    The first requirement translates into $\alpha_{ij}-\alpha_{i0}\leq t_{ij}\leq \gamma_{ij}-\gamma_{0j}$ for all matched $(i,j)$, that is if $d_{ij}=1$. The second one is easier to spell out if we define $u_i$ (resp.\ $v_j$) to be the post-transfer utility of man $i$ (resp.\ woman $j$) at the stable matching. Then we require that if $d_{ij}=0$,  we cannot find a value of the transfer $t_{ij}$ that satisfies both $\alpha_{ij}+t_{ij} > u_i$ and $\gamma_{ij}-t_{ij} > v_j$.   Obviously, this is equivalent to requiring that $\tilde{\Phi}_{ij} \leq u_i+v_j$. Note that if $d_{ij}=1$, then this inequality is binding since the joint surplus must be the sum of the post-transfer utilities. Moreover, the first requirement can be rewritten as $u_i \geq \alpha_{i0}$ for all  men and $v_j\geq  \gamma_{0j}$, with equality if man $i$ or woman $j$ is single.

    We summarize this in an equivalent definition of \idf{stability}{Stability}.
    \begin{definition}[Stability---dual definition]\label{ch:structuraltumodels:def:stability:ptu2}
        A feasible matching  $\bm{d}$ is stable if and only if the post-transfer utilities $u_i$ and $v_j$ satisfy
        \begin{itemize}
            \item for all $i$, $u_i \geq \tilde{\Phi}_{i0}$, with equality if $i$ is unmatched; and for all $j$, $v_j \geq \tilde{\Phi}_{0j}$, with equality if $j$ is unmatched
            \item for all $i$ and $j$, $u_i + v_j \geq \tilde{\Phi}_{ij}$, with equality if $i$ and $j$ are matched.
        \end{itemize}
    \end{definition}
 The conditions in Definition~\ref{ch:structuraltumodels:def:stability:ptu2} are exactly  the \idf{Karush-Kuhn-Tucker optimality conditions}{Karush-Kuhn-Tucker optimality conditions} of the following maximization program:
\begin{eqnarray}
\max_{\bm{d}\geq 0}%
  &&\sum_{i, j} d_{ij} \tilde{\Phi}_{ij} +\sum_i d_{i0}\tilde{\Phi}_{i0}+
  \sum_j d_{0j}\tilde{\Phi}_{0j} \nonumber\\
    s.t.~ &&\sum_j d_{ij} +d_{i0}= 1\;\;\forall i \\
    &&\sum_i d_{ij}+d_{0j}=1\;\;\forall j \; \; \; \; \label{ch:structuraltumodels:pgm:primal}
    \end{eqnarray}
    if $u_i$ and $v_j$ are the multipliers of the feasibility conditions. Thus the stable matchings maximize the total \idf{joint surplus}{Joint surplus} under the feasibility constraints. Program~\ref{ch:structuraltumodels:pgm:primal} above is called the \idf{{\em primal program}}{Primal program}. Since both the objective function and the constraints are linear, its  {\em dual\/} has the same value. It minimizes the sum of the post-transfer utilities under the stability constraints
    \begin{eqnarray}
        \min_{(u_i), (v_j)}%
          &&\sum_i u_i +\sum_j v_j \nonumber\\
            s.t.~  && u_i \geq \tilde{\Phi}_{i0}\;\;\forall i\nonumber\\
            && v_j \geq \tilde{\Phi}_{0j}\;\;\forall j\nonumber\\
            &&u_i+v_j \geq \tilde{\Phi}_{ij}\;\;\forall i, j, \; \; \; \; \label{ch:structuraltumodels:pgm:dual}
            \end{eqnarray}
            and the multipliers of the  constraints equal  the $d_{i0}, d_{0j}, d_{ij}$ of the associated stable matching.

From an economic point of view, the linearity of these programs implies that since the feasibility set is never empty (one can always leave all men and women unmatched), there exists a stable matching, it is generically unique, and there always exists a stable matching $\bm{d}$ whose  elements are all integers (zero or one). This paints a very different picture from matching with non-transferable utility.

\subsection{Separability}
A proper econometric setting requires that we  distinguish carefully what the analyst can observe from
\idf{{\em unobserved heterogeneity}}{Unobserved heterogeneity}, which only the market participants observe. Most crucially, the analyst cannot observe all the
determinants of the pre-transfer utilities $\alpha_{ij}$  and $\gamma_{ij}$ generated by a
hypothetical match between a man $i$ and a woman $j$. A priori, they could
depend on interactions between  characteristics the analyst observes, between these characteristics and unobserved
heterogeneity, and between the unobserved heterogeneity of both partners.

We now define observed characteristics as  {\em types \/} $x\in \mathcal{X}$ for men, and $y\in \mathcal{Y}$ for women. These \idf{types}{Types} are observed by all market participants as well as the analyst.
There are $n_{x}$ men of type $x$ and $m_{y}$ women of type $y$. The set of
marital options that are offered to men and women is the set of types of
partners on the other side of the market, plus singlehood. We continue to use the
notation $0$ for singlehood and we define $\mathcal{X}_{0}=\mathcal{X}\cup
\{0\}$ and $\mathcal{Y}_{0}=\mathcal{Y}\cup \{0\}$ as the set of
options that are available to respectively women and men.

Men and women of a given type also have other characteristics which are not observed by the analyst. A man $i$ who has observed type $x$, or a woman $j$ who has observed type $y$, may be a more or less appealing partner in any number of ways. In so far as these characteristics are payoff-relevant, they contribute to determining who matches whom. We will assume in this chapter that contrary to the analyst, all participants observe these additional characteristics.  To the analyst, they constitute {\em unobserved heterogeneity}.  It is important to note that this distinction is data-driven: richer data converts \idf{unobserved heterogeneity}{Unobserved heterogeneity} into types.

Much of the literature has settled on excluding interactions between
unobserved characteristics, and this is the path we take here. We impose:
\begin{assumption}[Separability]\label{ch:structuraltumodels:assn:separ}
The joint surplus generated by a match between man $i$ with type $x$ and woman $j$ with type $y$ is
\begin{equation}
    \label{ch:structuraltumodels:eq:separ}
    \tilde{\Phi}_{ij} = \Phi_{xy} + \varepsilon_{iy} + \eta_{jx}.
\end{equation}
The utility of
man $i$ and woman $j$ if unmatched are $\varepsilon _{i0}$ and $\eta _{j0}$
respectively.
 \end{assumption}

In the language of analysis of variance models, the \idf{separability}{Separability} assumption
rules out two-way interactions between unobserved characteristics,
conditional on observed \idf{types}{Types}. While this is restrictive, it still allows
for rich patterns of matching in equilibrium. For instance, all women may
like educated men, but those women who give a higher value to education are more likely
(everything equal) to marry a more educated man, provided that they in turn
have observed or unobserved characteristics that more educated men value
more.

Since  the analyst can only observe types, we now redefine a matching as a collection $\bm{\mu}$ of non-negative numbers:
 $\mu_{xy}$ denotes the number of matches between men of type $x$ and
women of type $y$, which is determined in equilibrium and observed by the analyst. All men of type $x$,
and all women of type $y$, must be single or matched. This generates the
feasibility constraints:
\begin{align*}
N_{x}(\bm{\mu})& :=\sum_{y\in \mathcal{Y}}\mu _{xy}+\mu _{x0}=n_{x}\;\;\forall
x\in \mathcal{X} \\
M_{y}(\bm{\mu})& :=\sum_{x\in \mathcal{X}}\mu _{xy}+\mu _{0y}=m_{y}\;\;\forall
y\in \mathcal{Y}.
\end{align*}

In the following, we denote  $x_i=x$ if man $i$ is of type $x$, and $y_j=y$ if woman $j$ is of type $y$.

\subsection{Equilibrium}
 \idf{Convex duality}{Convex duality} will be the key to our approach to
\idf{identification}{Identification}.  We start by rewriting the dual characterization of
the stable matching in~\eqref{ch:structuraltumodels:pgm:dual} as
\begin{eqnarray}
\min_{\substack{ u_{i}\geq \varepsilon_{i0}  \\ v_{j}\geq \eta_{j0}}}
&&\left(\sum_{i} u_i+\sum_{j} v_j\right)  \label{ch:structuraltumodels:eq:dualgal} \\
s.t.~ &&u_{i}+v_{j}\geq \tilde{\Phi}_{ij} \; \; \forall i,j.  \notag
\end{eqnarray}
Given Assumption~\ref{ch:structuraltumodels:assn:separ}, the constraint in~\eqref{ch:structuraltumodels:eq:dualgal} can
be rewritten as
\begin{equation}
(u_i -\varepsilon_{iy})+(v_j -\eta_{jx}) \geq \Phi_{x_i y_j} \; \; \forall
i, j.  \label{ch:structuraltumodels:eq:keysep}
\end{equation}
Define $U_{xy}=\min_{i: x_{i}=x}\left\{ u_{i}-\varepsilon _{iy}\right\} $
and $V_{xy}=\min_{j: y_{j}=y}\left\{ v_{j}-\eta _{jx}\right\}$ for $x,y\neq 0$; and without loss of generality, set $U_{x0}=V_{0y}=0$ for $x,y>0$. The
constraint becomes
\begin{equation*}
U_{xy}+V_{xy} \geq \Phi_{xy} \; \; \forall x,y.
\end{equation*}
Moreover, by definition $u_i = \max_{y\in \mathcal{Y}_0}(U_{x_i
y}+\varepsilon_{iy})$ and $v_j = \max_{x\in \mathcal{X}_0}(V_{x
y_j}+\eta_{jx})$, so that we can rewrite the \idf{dual program}{Dual program} as
\begin{eqnarray*}
\min_{\bm{U}, \bm{V}} &&\left(\sum_{i} \max_{y\in \mathcal{Y}_0}(U_{x_i
y}+\varepsilon_{iy}) + \sum_j \max_{x\in \mathcal{X}_0}(V_{x
y_j}+\eta_{jx})\right) \\
s.t.~ &&U_{xy}+V_{xy}\geq \Phi _{xy} \; \; \forall x,y.
\end{eqnarray*}

Inspection of the objective function shows that the inequality constraint $U_{xy}+V_{xy}\geq
\Phi _{xy}$ can be replaced by an equality; indeed, if it were strict, one could weakly improve the objective function
while satisfying the constraint. Since this implies that $U_{xy}+V_{xy}=\Phi_{xy}$,
we can replace $V_{xy}$ with $%
(\Phi_{xy}-U_{xy})$ to obtain a simple formula for the total \idf{joint surplus}{Joint surplus}:
\begin{equation}
    \label{ch:structuraltumodels:eq:minmax}
  \mathcal{W} =
\min_{\bm{U}} \left(\sum_i \max_{y\in \mathcal{Y}_0}(U_{x_i y}+\varepsilon_{iy}) +
\sum_j \max_{x\in \mathcal{X}_0}(\Phi_{xy_j}-U_{x y_j}+\eta_{jx})\right).
\end{equation}%
We just reduced the dimensionality of the problem from the number of
individuals in the market to the product of the numbers of their observed
types. Since the latter is typically orders of magnitude smaller than the
former, this is a drastic simplification. Assumption~\ref{ch:structuraltumodels:assn:separ} was
the key ingredient: without it, we would have an unobserved  term $\xi_{ij}$ interacting
the unobservables in the joint surplus $\tilde{\Phi}_{ij}$ and~%
\eqref{ch:structuraltumodels:eq:keysep} would lose its nice separable structure.

Moreover, the nested min-max in equation~\eqref{ch:structuraltumodels:eq:minmax} is not as complex as it seems. Consider the expression
\[
 G_x(U_{x\cdot}):=
\frac{1}{n_x} \sum_{x_i = x} \max_{y\in \mathcal{Y}_0}(U_{x y}+\varepsilon_{iy}).
\]
 When the number  of individuals $n_x$ tends to infinity, $G_x$ converges  to the Emax operator, namely
\[
 G_x(U_{x\cdot}):=
\mathbb{E} [ \max_{y\in \mathcal{Y}_0}(U_{x y}+\varepsilon_{iy}) ].
\]
We shall assume   from now on that this {\em large  market limit\/} is a good approximation.

Since the maximum is taken over a collection of linear functions of $\bm{U}_{x\cdot}$, its value is a convex function, and so is $G_x$.   Defining $H_y(\bm{V}_{\cdot y})$ similarly, we obtain
\begin{equation}
\label{ch:structuraltumodels:eq:WGH}
\mathcal{W} =     \min_{\bm{U}} \left(G(\bm{U})+H(\bm{\Phi}-\bm{U})\right)
\end{equation}
where
\begin{align*}
    G(\bm{U}) &:=  \sum_{x\in \mathcal{X}} n_x G_x(\bm{U}_{x\cdot})\\
    H(\bm{V}) &:=  \sum_{y\in \mathcal{Y}} m_y H_y(\bm{V}_{\cdot y}).
\end{align*}
These functions play a special role in our analysis. Since $G$ is convex, it has a subgradient everywhere, which is a singleton almost everywhere. It is easy to see that the derivative of $\max_{y\in \mathcal{Y}_0}(U_{x y}+\varepsilon_{iy})$ with respect to $U_{xy}$ equals~1 if $y$ achieves a strict maximum, and~0 if it is not a maximum.  As a consequence, the subgradient of $G_x$ with respect to $U_{xy}$ is\footnote{Neglecting the measure zero cases where the subgradient is not a singleton.}  the proportion of men of type $x$ whose match is of type $y$. We denote this proportion $\mu^M_{y\vert x}$. Finally, we note that the subgradient of $G$ with respect to $U_{xy}$ is $n_x$ times the subgradient of $G_x$, that is the number $\mu^M_{xy}$ . To conclude (and using similar definitions for $H$):
\begin{align*}
    \bm{\mu}^M  = \partial G(\bm{U})\\
    \bm{\mu}^W  = \partial H(\bm{V}).
\end{align*}
In equilibrium we must have $\mu^M_{xy}=\mu^W_{xy}$ for all $x,y$.  This should not come as a surprise as it translates the first-order conditions in~\eqref{ch:structuraltumodels:eq:WGH}:
\[
    \partial G(\bm{U})\cap \partial H(\bm{\Phi}-\bm{U}) \neq \emptyset.
    \]

\section{Identification}\label{ch:structuraltumodels:sec:ident}
Now let us denote $%
G^{\ast }$ the \idf{Legendre-Fenchel transform}{Legendre-Fenchel transform} of the convex function $G$:
\begin{equation*}
G^{\ast }(\mu )=\sup_{\bm{a}}\left( \sum
_{\substack{ x\in \mathcal{X}  \\ y\in \mathcal{Y}}}
\mu_{xy}a_{xy}-G(a)\right) .
\end{equation*}%
It is another convex function; and by the theory of
\idf{convex
duality}{convex duality} we know that since
\begin{equation*}
\bm{\mu}^{M}= \partial G(\bm{U}),
\end{equation*}%
we also have $\bm{U} =  G^{\ast }(\bm{\mu}^{M})$, that is
\begin{equation}
U_{xy} = \frac{\partial G^\ast}{\partial\mu_{xy}}(\bm{\mu}^{M}). \label{ch:structuraltumodels:eq:ident:x:types}
\end{equation}
Similarly,
\begin{equation}
    V_{xy} = \frac{\partial H^\ast}{\partial\mu_{xy}}(\bm{\mu}^{W}). \label{ch:structuraltumodels:eq:ident:y:types}
    \end{equation}

\subsection{Identifying the Joint Surplus}
In equilibrium, $\bm{\mu}^M=\bm{\mu}^W:=\bm{\mu}$ and $\bm{U}+\bm{V}=\bm{\Phi}$; therefore we obtain
\begin{equation}
    \Phi_{xy} = \frac{\partial G^\ast}{\partial \mu_{xy}}(\bm{\mu})+\frac{\partial H^\ast}{\partial \mu_{xy}}(\bm{\mu}). \label{ch:structuraltumodels:eq:ident:Phi}
\end{equation}
Observing the matching patterns thus identifies
all values of $U_{xy}$, $V_{xy}$, and $\Phi_{xy}$, provided that we have enough information to evaluate the function $G$. Since the shape of the function $G$ only depends on the distribution of the \idf{unobserved heterogeneity}{Unobserved heterogeneity} terms, this is the piece of information we need.

\begin{assumption}[Distribution of the unobserved heterogeneity]\label{ch:structuraltumodels:assn:distrib_unobs}
    For any man $i$ of type $x$, the random vector $\bm{\varepsilon} _{i\cdot}={%
    (\varepsilon _{iy})}_{y\in \mathcal{Y}_{0}}$ is distributed according to $%
    \mathbb{P}_{x}$.

    Similarly, for any woman $j$ of type $y$, the random vector
    $\bm{\eta} _{j\cdot}={(\eta _{jx})}_{x\in \mathcal{X}_{0}}$ is distributed according
    to $\mathbb{Q}_{y}$.
\end{assumption}

Note that $\eqref{ch:structuraltumodels:eq:ident:Phi}$ is a system of $%
\abs{\mathcal{X}}\times \abs{\mathcal{Y}}$ equations. To repeat, it
identifies the $\bm{\Phi}$ matrix in the joint surplus as a function of the
observed matching patterns $(\bm{\mu})$ and the shape of the functions $%
G^{\ast }$ and $H^{\ast }$. The latter in turn only depend on the distributions $%
\mathbb{P}_{x}$ and $\mathbb{Q}_{y}$. It is important to stress that the joint surplus is uniquely identified given any choice of these distributions. Identifying the distributions themselves requires more restrictions and/or more data.

\subsection{Generalized Entropy}
We already know from Section~\ref{ch:structuraltumodels:sub:stabil} that the stable matching maximizes the
total joint surplus. The corresponding \idf{primal program}{Primal program} is
\begin{equation}
\mathcal{W}(\bm{\Phi},\bm{n},\bm{m})=\max_{\bm{\mu} \geq 0}\left( \sum_{\substack{ x\in
\mathcal{X}  \\ y\in \mathcal{Y}}}\mu _{xy}\Phi _{xy}-\mathcal{E}\left(\bm{\mu}; \bm{n},\bm{m}\right) \right)  \label{ch:structuraltumodels:eq:primal:gal}
\end{equation}%
where
\begin{equation*}
\mathcal{E}\left(\bm{\mu}; \bm{n},\bm{m}\right) =G^{\ast}\left(\bm{\mu}; \bm{n}\right) +H^{\ast
}\left(\bm{\mu},\bm{m}\right)
\end{equation*}%
is the \idf{generalized entropy}{generalized entropy} of the matching $\bm{\mu}$.  It is easy to check that the first-order
conditions in~\eqref{ch:structuraltumodels:eq:primal:gal} (which is globally concave) coincide
with the \idf{identification}{Identification} formula~\eqref{ch:structuraltumodels:eq:ident:Phi}.

The two parts of the objective function in~\eqref{ch:structuraltumodels:eq:primal:gal} have a
natural interpretation. The sum $\sum_{x,y}\mu_{xy}\Phi_{xy}$ reflects the
value of matching on observed \idf{types}{Types} only. The generalized entropy term $-%
\mathcal{E}(\bm{\mu}; \bm{n},\bm{m})$ is the sum of the values that are generated by
matching unobserved heterogeneities with observed types: e.g.\ men of type $%
x $ with a high value of $\varepsilon_{iy}$ being more likely to match with
women of type $y$.

We skipped over an important technical issue: the
Legendre-Fenchel transform of $G_x$ is equal to $+\infty $ unless $%
\sum_{y\in \mathcal{Y}}\mu _{xy}=N_x(\bm{\mu})-\mu_{x0}\leq n_{x}$. Therefore the
objective function in~\eqref{ch:structuraltumodels:eq:primal:gal} is minus infinity when any of
these feasibility constraints is violated.
There are two approaches for making the problem well-behaved. We can simply
add the constraints to the program. As it turns out, extending the
generalized entropy beyond its domain is sometimes a much better approach, as we will show in Section~\ref{ch:structuraltumodels:sec:estimation}.

\subsection{The Logit Model\label{ch:structuraltumodels:sub:mlogit}}

Following a long tradition in discrete choice models, much of the literature
has focused on the case when the distributions $\mathbb{P}_{x}$ and $\mathbb{%
Q}_{y}$ are \idf{standard type~I extreme value (Gumbel)}{Gumbel distribution}. Under this
distributional assumption, the $G_{x}$ functions take a very simple and
familiar form:
\begin{equation*}
G_{x}(U_{x\cdot })=\log\left(1+\sum_{t\in \mathcal{Y}}\exp
(U_{xt})\right);
\end{equation*}%
and the generalized entropy function $\mathcal{E}$ is just the usual
\idf{entropy}{Entropy}:
\begin{equation}
\mathcal{E}\left( \bm{\mu}; \bm{n}, \bm{m} \right) =2\sum_{\substack{ x\in \mathcal{X}  \\ %
y\in \mathcal{Y}}}\mu _{xy}\log \mu _{xy}+\sum_{x\in \mathcal{X}}\mu
_{x0}\log \mu _{x0}+\sum_{y\in \mathcal{Y}}\mu _{0y}\log \mu _{0y}.
\label{ch:structuraltumodels:EisEntropy}
\end{equation}%
Equation~\eqref{ch:structuraltumodels:eq:ident:Phi} can be rewritten to yield the following %
\idf{matching function}{Matching function}, which links the numbers of
singles, the joint surplus, and the numbers of matches:
\begin{equation}
\mu _{xy}=\sqrt{\mu _{x0}\mu _{0y}}\exp \left( \frac{\Phi _{xy}}{2}\right).
\label{ch:structuraltumodels:eq:csmmf}
\end{equation}%

In the  \idf{logit}{Logit} model, the distributions $\mathbb{P}_x$ and $%
\mathbb{Q}_y$ have no free parameter: the only unknown parameters in the
model are those that determine the joint surplus matrix $\Phi$. Using~%
\eqref{ch:structuraltumodels:eq:csmmf} gives %
\idf{Choo and Siow's formula}{Choo and Siow's formula}:
\begin{equation}
\Phi_{xy}=\log \frac{\mu_{xy}^{2}}{\mu_{x0}\mu_{0y}}  \label{ch:structuraltumodels:eq:csident}
\end{equation}

\section{Estimation}\label{ch:structuraltumodels:sec:estimation}

In matching markets, the sample may be drawn from the population at the
individual level or at the match level. Take the marriage market as an
example. With individual sampling, each man or woman in the population would
be a sampling unit. In fact, household-based sampling is more common in
population surveys: when a household is sampled, data is collected on all of
its members. Some of these households consist of a single man or woman, and
others consist of a married couple. We assume here that sampling is at the
household level.

Recall that $\hat{\mu}_{xy}$, $\hat{\mu}_{x0}$ and $\hat{\mu}_{0y}$ are the
number of matches of type $(x,y)$, $(x,0)$ and $(0,y)$, respectively in our sample.
Denote
\begin{equation*}
N_{h}=\sum_{x\in \mathcal{X}}\hat{\mu}_{x0}+\sum_{y\in \mathcal{Y}}\hat{\mu}%
_{0y}+\sum_{\substack{ x\in \mathcal{X}  \\ y\in \mathcal{Y}}}\hat{\mu}_{xy}
\end{equation*}%
the number of households in our sample, and let
\begin{equation*}
\hat{\bm{\pi}}_{xy}=\frac{\hat{\mu}_{xy}}{N_{h}},\hat{\pi}_{x0}=\frac{\hat{\mu}%
_{x0}}{N_{h}}\text{ and }\hat{\bm{\pi}}_{0y}=\frac{\hat{\mu}_{0y}}{N_{h}}
\end{equation*}%
the empirical sample frequencies of matches of type $(x,y)$, $(x,0)$ and $(0,y)$,
respectively. Let $\bm{\pi}$ be the population analog of $\hat{\bm{\pi}}$.
The estimators of the matching probabilities have an asymptotic distribution
\begin{equation}\label{ch:structuraltumodels:eq:Vpi}
\hat{\bm{\pi}}\sim \mathcal{N}\left( 0,\frac{\bm{V_{\pi}}}{N_{h}}\right).
\end{equation}%

We seek to estimate a \idf{parametric model}{Parametric model} of the matching market. This involves specifying functional form for the matrix $\bm{\Phi}$ and choosing families of distributions for the \idf{unobserved heterogeneity}{Unobserved heterogeneity} $\mathbb{P}_x$ and $\mathbb{Q}_y$. We denote $\bm{\lambda}$ the parameters of $\bm{\Phi}$, $\bm{\beta}$ the parameters of the distributions, and  our aim is  to estimate $\bm{\theta}=(\bm{\lambda},\bm{\beta})$. Depending on the context, the analyst may choose to allocate more
parameters to the matrix $\Phi$ or to the distributions $\mathbb{P}_x$ and $%
\mathbb{Q}_y$. We assume that the model is well-specified in that the data was generated by a matching market with true parameters $\bm{\theta}_0$.

We will assume in this section that the analyst is able to compute the stable matching $\bm{\mu}^{\bm{\theta}}$ for any value of the parameters $\bm{\theta}$. We provide several ways to do so efficiently in Section~\ref{ch:structuraltumodels:sec:computation}.

\subsection{The Maximum Likelihood Estimator}
In this setting, the log-likelihood function of the sample is simply the sum over all households of the log-probabilities of the observed matches. Let us fix the  value of the parameters of the model at $\bm{\theta}$. We denote $\bm{\mu}^{\bm{\theta}}$ the equilibrium matching patterns for these values of the parameters and the observed margins $\bm{n}$ and $\bm{m}$.

A household may consist of a match between a man of type $x$ and a woman of type $y$, of a single man of type $x$, or of a single woman of type $y$. The corresponding probabilities are respectively $\mu^{\bm{\theta}}_{xy}/N_h^{\bm{\theta}}$, $\mu^{\bm{\theta}}_{x0}/N_h^{\bm{\theta}}$, and $\mu^{\bm{\theta}}_{0y}/N_h^{\bm{\theta}}$, where
\[
N_h^{\bm{\theta}}:= \sum_{x,y \in \mathcal{X}\times \mathcal{Y}} \mu^{\bm{\theta}}_{xy} + \sum_{x\in \mathcal{X}} \mu^{\bm{\theta}}_{x0}+\sum_{y \in\mathcal{Y}}\mu^{\bm{\theta}}_{0y}
\]
is the number of households in the stable matching for $\bm{\theta}$, which in general differs from $N_h$. The log-likelihood becomes
\begin{align*}\label{ch:structuraltumodels:eq:loglike}
    \log L(\bm{\theta}) :=    \sum_{x,y \in \mathcal{X}\times \mathcal{Y}} \hat{\mu}_{xy} \log \frac{\mu^{\bm{\theta}}_{xy}}{N_h^{\bm{\theta}}} + \sum_{x\in \mathcal{X}} \hat{\mu}_{x0} \log \frac{\mu^{\bm{\theta}}_{x0}}{N_h^{\bm{\theta}}}
    +\sum_{y\in \mathcal{Y}}  \hat{\mu}_{0y} \log \frac{\mu^{\bm{\theta}}_{0y}}{N_h^{\bm{\theta}}}.
\end{align*}
Maximizing this expression gives a \idf{maximum likelihood estimator}{Maximum likelihood estimator} that has the usual asymptotic properties: it is consistent, asymptotically normal, and asymptotically efficient. The maximization process may not be easy, however. In particular, the function $\log L$ is unlikely to be globally concave, and it may have several local extrema. This may make other approaches more attractive.

\subsection{The Moment Matching Estimator}\label{ch:structuraltumodels:sub:momatch}
A natural choice of parameterization for $\bm{\Phi}^{\bm{\lambda}}$ is the linear
expansion
\begin{equation*}
\Phi _{xy}^{\bm{\lambda}}=\sum_{k=1}^{K}\lambda _{k}\phi_{xy}^{k}
\end{equation*}%
where the basis functions $\phi _{xy}^{k}$ are given and the $\lambda_{k}$
coefficients are to be estimated.

The \idf{moment matching estimator}{Moment-matching estimator} uses the $K$ equalities
\[
    \sum_{x,y} \mu^{\bm{\theta}}_{xy} \phi^k_{xy}= \sum_{x,y} \hat{\mu}_{xy} \phi^k_{xy}
    \]
    as its estimating equations. Both sides of these equalities can be interpreted as expected values of the basis function $\phi^k$; in this sense, the estimator matches the observed and simulated (first)  moments of the basis functions.  By construction, it can only identify $K$ parameters. We assume from now on that the values of the parameters of the distribution are fixed at $\bm{\beta}$, and we seek to estimate $\bm{\lambda}$.

Applying the envelope theorem to equation~\eqref{ch:structuraltumodels:eq:primal:gal} shows that the derivative of the total joint surplus with respect to $\Phi_{xy}$ is the value of $\mu_{xy}$ for the corresponding stable matching. Using the chain rule, we obtain
\[
\frac{\partial \mathcal{W}^{\bm{\beta}}}{\partial\lambda_k}(\bm{\mu}^{\bm{\theta}}, \hat{\bm{n}}, \hat{\bm{m}})= \sum_{x,y} \mu^{\bm{\theta}}_{xy}\phi^k_{xy};
\]
this allows us to rewrite the moment matching estimating equations as the first order conditions of
\begin{equation}\label{ch:structuraltumodels:eq:mom:match}
    \max_{\bm{\lambda}}\left(\sum_{x,y} \hat{\mu}_{xy}\Phi^{\bm{\lambda}}_{xy}
    -\mathcal{W}^{\bm{\beta}}(\bm{\mu}^{\bm{\theta}}, \hat{\bm{n}}, \hat{\bm{m}})\right).
\end{equation}
    Note that the function $\mathcal{W}$ is convex in $\bm{\Phi}$. Since $\bm{\Phi}^{\bm{\lambda}}$ is linear in $\bm{\lambda}$, the objective function of~\eqref{ch:structuraltumodels:eq:mom:match} is globally convex. This is of course a very appealing property in a maximization problem.

    We still have to evaluate $\mathcal{W}^{\bm{\beta}}(\bm{\mu}^{\bm{\theta}}, \hat{\bm{n}}, \hat{\bm{m}})=\sum_{x,y} \mu^{\bm{\theta}}_{xy} \Phi^{\bm{\lambda}}_{xy} -\mathcal{E}^{\bm{\beta}}(\bm{\mu}^{\bm{\theta}}; \hat{\bm{n}}, \hat{\bm{m}})$. It is often possible to circumvent that step, however. To see this, remember that the \idf{generalized entropy}{Generalized entropy} is only defined when $\bm{N}(\bm{\mu})=\hat{\bm{n}}$ and $\bm{M}(\bm{\mu})=\hat{\bm{m}}$. Now take any real-valued functions $f$ and $g$ such that $f(0)=g(0)=0$, and consider the {\em extended entropy\/} function
\[
    E^{\bm{\beta}}(\bm{\mu}; \hat{\bm{n}},\hat{\bm{m}}) = \mathcal{E}^{\bm{\beta}}(\bm{\mu}; \bm{N}(\bm{\mu}),\bm{M}(\bm{\mu}))
    +f(\bm{N}(\bm{\mu})-\hat{\bm{n}})+g(\bm{M}(\bm{\mu})-\hat{\bm{m}}).
    \]
    By construction, this function is well-defined for any $\bm{\mu}$, and it coincides with $\mathcal{E}^{\bm{\beta}}$ when $\bm{N}(\bm{\mu})=\hat{\bm{n}}$ and $\bm{M}(\bm{\mu})=\hat{\bm{m}}$. Therefore we can rewrite~\eqref{ch:structuraltumodels:eq:primal:gal} as
    \begin{eqnarray*}
    \mathcal{W}^{\bm{\beta}}(\bm{\Phi},\bm{n},\bm{m})=\max_{\bm{\mu} \geq 0}\left( \sum_{\substack{ x\in
    \mathcal{X}  \\ y\in \mathcal{Y}}}\mu _{xy}\Phi _{xy}-E^{\bm{\beta}}\left(\bm{\mu}; \bm{n},\bm{m}\right) \right)  \\
    s.t \; \bm{N}(\bm{\mu})=\hat{\bm{n}}  \; \mbox{ and } \; \bm{M}(\bm{\mu})=\hat{\bm{m}}.
    \end{eqnarray*}%
If moreover we choose $f$ and $g$ to be convex functions, this new program is also convex. As such,  it has a dual formulation that can be written in terms
of the Legendre-Fenchel transform $(E^{\bm{\beta}})^\ast$ of $E^{\bm{\beta}}$.  Simple
calculations show that the dual is:
\begin{equation*}
\mathcal{W}^{\bm{\beta}}\left(\bm{\Phi},\hat{\bm{n}},\hat{\bm{m}}\right) =\min_{\bm{u},\bm{v}\geq 0}\left(\dotprod{\hat{\bm{n}}}{\bm{u}}+%
\dotprod{\hat{\bm{m}}}{\bm{v}}+(E^{\bm{\beta}})^\ast\left(\bm{\Phi} -\bm{u}-\bm{v},-\bm{u},-\bm{v}\right)\right)
\end{equation*}%
where we denote $\bm{\Phi} -\bm{u}-\bm{v}={(\Phi _{xy}-u_{x}-v_{y})}_{x,y}$.

Returning to~\eqref{ch:structuraltumodels:eq:mom:match}, the program that defines the moment matching estimator can now be rewritten as follows:
\begin{equation}\label{ch:structuraltumodels:eq:mm:Estar}
    \max_{\bm{\lambda}, \bm{u}\geq 0, \bm{v}\geq 0}\left(\sum_{x,y} \hat{\mu}_{xy}\Phi^{\bm{\lambda}}_{xy}
    -\dotprod{\hat{\bm{n}}}{\bm{u}}
    -\dotprod{\hat{\bm{m}}}{\bm{v}}-(E^{\bm{\beta}})^\ast\left(\bm{\Phi} -\bm{u}-\bm{v},-\bm{u},-\bm{v}\right)\right).
\end{equation}
This is still a globally convex program; and if we can choose $f$ and $g$ such that the  extended entropy $(E^{\bm{\beta}})^\ast$ has a simple Legendre-Fenchel transform, it will serve as  a computationally
attractive estimation procedure.
 In addition to estimating the parameters  $\bm{\lambda}$ of the \idf{joint surplus}{Joint surplus}, it
directly yields estimates of the expected utilities $\bm{u}$ and $\bm{v}$ of each type. Moreover, after estimation the matching patterns can be obtained by:
\begin{equation}
    \left\{
    \begin{array}{c}
    \mu^{\bm{\theta}} _{xy}=\frac{\partial (E^{\bm{\beta}})^\ast}{\partial z_{xy}}\left(\bm{\Phi}-\bm{u}-\bm{v},-\bm{u},-\bm{v}%
    \right) \\[3mm]
    \mu^{\bm{\theta}} _{x0}=\frac{\partial (E^{\bm{\beta}})^\ast}{\partial z_{x0}}\left(\bm{\Phi}-\bm{u}-\bm{v},-\bm{u},-\bm{v}\right) \\[3mm]
    \mu^{\bm{\theta}} _{0y}=\frac{\partial (E^{\bm{\beta}})^\ast}{\partial z_{0y}}\left(\bm{\Phi}-\bm{u}-\bm{v},-\bm{u},-\bm{v}\right)%
    \end{array}%
    \right.  \label{ch:structuraltumodels:def_pi_theta}
    \end{equation}
The logit model of Section~\ref{ch:structuraltumodels:sub:mlogit} provides an illustration of this approach.

\subsection{Estimating the Logit Model}\label{ch:structuraltumodels:sub:logitEstim}
Plugging in estimates $\hat{\bm{\mu}}$ of the matching patterns in  formula~\eqref{ch:structuraltumodels:eq:csident}
gives a closed-form estimator $\hat{\bm{\Phi}}$ of the \idf{joint surplus}{Joint surplus} matrix in the \idf{logit model}{Logit}. On
the other hand, determining the equilibrium matching patterns $\bm{\mu}$ for
given primitive parameters $\bm{\Phi}, \bm{n}, \bm{m}$ is more involved; and it is
necessary in order to evaluate counterfactuals that modify these primitives
of the model. We will show how to do it in Section~\ref{ch:structuraltumodels:sub:ipfp} below. In
addition, the analyst may want to assume that the joint surplus
matrix $\bm{\Phi}$ belongs in a parametric family $\bm{\Phi}^{\lambda}$. While this
could be done by finding the value of $\bm{\lambda}$ that minimize the distance
between $\bm{\Phi}^{\lambda}$ and the $\hat{\bm{\Phi}}$ obtained from~%
\eqref{ch:structuraltumodels:eq:csident}, the approach sketched in Section~\ref{ch:structuraltumodels:sub:momatch}  is more appealing.

To construct an \idf{extended entropy}{Extended entropy} function $E$ in the logit model, we rely on the primitive of
the logarithm $\mathcal{L}(t)= t\log t-t$; we define $f(\bm{T})= \sum_{x} \mathcal{L}(T_x)$, and similarly for $g$. They are clearly convex functions.
The reason for this a priori non-obvious choice of strictly convex functions
is that many of the terms in the derivatives of the resulting extended
entropy  cancel out. In fact, simple calculations give
\begin{equation}
E^{\ast }\left(\bm{z}\right) =2 \sum_{\substack{ x\in \mathcal{X}  \\ y\in
\mathcal{Y}}}\exp \left(\frac{z_{xy}}{2}\right) + \sum_{x\in \mathcal{X}%
}\exp \left(z_{x0}\right) + \sum_{y\in \mathcal{Y}}\exp \left(z_{0y}\right).
\label{ch:structuraltumodels:Estar-logit}
\end{equation}%
Substituting into~\eqref{ch:structuraltumodels:eq:mm:Estar}, the moment matching estimator and associated utilities solve
\begin{equation*}
\min_{\bm{\lambda}, \bm{u}\geq 0, \bm{v}\geq 0}F\left(\bm{\lambda},\bm{u}, \bm{v}\right)
\end{equation*}%
where
\begin{eqnarray*}
F\left(\bm{\lambda},\bm{u}, \bm{v}\right) &=&\sum_{x\in \mathcal{X}}\exp \left(
-u_{x}\right) +\sum_{y\in \mathcal{Y}}\exp \left( -v_{y}\right) +2\sum
_{\substack{ x\in \mathcal{X}  \\ y\in \mathcal{Y}}}\exp \left( \frac{\Phi
_{xy}^{\lambda }-u_{x}-v_{y}}{2}\right) \\
&&-\sum_{\substack{ x\in \mathcal{X}  \\ y\in \mathcal{Y}}}\hat{\pi}%
_{xy}\left( \Phi _{xy}^{\lambda }-u_{x}-v_{y}\right) +\sum_{x\in \mathcal{X}}%
\hat{\pi}_{x0}u_{x}+\sum_{y\in \mathcal{Y}}\hat{\pi}_{0y}v_{y}.
\end{eqnarray*}

 This is the objective function of a \idf{Poisson regression}{Poisson regression} with two-way fixed effects.
Minimizing $F$ is a very easy task; we give some specialized algorithms in Section~\ref{ch:structuraltumodels:sec:computation}, but problems of moderate size can also be treated using statistical packages handling generalized linear models. Denote $\bm{\alpha}=(\bm{\lambda},\bm{u}, \bm{v})$ the set of arguments of $F$.  The asymptotic distribution of the estimator of $\bm{\alpha}$ is given in Appendix~B.

\subsection{The Maximum-score Method}\label{ch:structuraltumodels:sub:max-score}

In most one-sided random utility models of discrete choice, the probability
that a given alternative is chosen increases with its mean utility. Assume
that alternative $k$ has utility $U(x_{kl},\theta_0)+u_{kl}$ for individual $%
l$. Let $K(l)$ be the choice of individual $l$ and for any given $\theta$,
denote
\begin{equation*}
R_l(\theta) \equiv \sum_{k\neq K(l)} \mathrm{1\kern-.40em 1}%
\left(U(x_{l,K(l)},\theta) > U(x_{kl},\theta)\right)
\end{equation*}
the rank (from the bottom) of the chosen alternative $K(l)$ among the mean
utilities. Choose any increasing function $F$. If (for simplicity) the $%
u_{kl}$ are iid across $k$ and $l$, maximizing the score function
\begin{equation*}
\sum_l F\left(R_l(\theta)\right)
\end{equation*}
over $\theta$ yields a consistent estimator of $\theta_0$. The underlying
intuition is simply that the probability that $k$ is chosen is an increasing
function of the differences of mean utilities $U(x_{kl},\theta) -
U(x_{k^\prime l},\theta)$ for all $k^\prime\neq k$.

It seems natural to ask whether a similar property also holds in two-sided
matching with transferable utility: is there a sense in which (under
appropriate assumptions) the probability of a match increases with the
surplus it generates?

If transfers are observed, then each individual's choices is just a
one-sided choice model and the maximum score estimator can be used
essentially as is. Without data on transfers, the answer is not
straightforward. In a two-sided model, the very choice of a single ranking
is not self-evident. In so far as the \idf{optimal matching}{Optimal matching} is partly driven by
unobservables, it is generally not true that the optimal matching maximizes
the joint total non-stochastic surplus for instance.

One can give a positive answer in one of the models we have already
discussed: the  \idf{logit}{Logit} specification of Section~\ref{ch:structuraltumodels:sub:mlogit}. Formula~\eqref{ch:structuraltumodels:eq:csmmf}
implies that for any $(x,x^\prime, y,y^\prime)$, the double log-odds ratio
$2 \log ( ( \mu_{xy} \mu_{x^\prime y^\prime} ) / ( \mu_{x,y^\prime} \mu_{x^\prime y} ))$
equals the double difference
\begin{equation*}
D_{\Phi}(x,x^\prime, y,y^\prime) \equiv \Phi_{xy} +
\Phi_{x^\prime y^\prime} - \Phi_{x^\prime y}-\Phi_{x y^\prime}.
\end{equation*}

This direct link between the observed matching patterns and the unknown
surplus function justifies a \idf{maximum-score estimator}{Maximum-score method}
\begin{equation*}
\max_{\bm{\Phi}} \sum_{(x,x^\prime, y,y^\prime) \in C} \mathrm{1\kern-.40em 1}%
\left(D_{\Phi}(x,x^\prime, y, y^\prime) > 0 \right)
\end{equation*}
where $C$ is a subset of the pairs that can be formed from the data.

More generally, one can prove the following result.

\begin{theorem}[Comonotonicity of double-differences]\label{ch:structuraltumodels:thm:maxscore:exch}
  Assume that the
    surplus is separable and that the distribution of the unobservable heterogeneity
    vectors is exchangeable.
    Then for all $(x,y,x^\prime,y^\prime)$, the log-odds
    ratio $ D_{\Phi}(x,x^\prime, y, y^\prime) $ and the double difference $ \log ( ( \mu_{xy} \mu_{x^\prime y^\prime} ) / ( \mu_{x,y^\prime} \mu_{x^\prime y} ))$ have the same sign.
\end{theorem}

 While this is clearly a weaker result than in the
\idf{logit}{Logit} model, it is enough to apply the same maximum-score
estimator.

One of the main advantages of the maximum-score method is that  it
extends to more complex matching markets. It also allows the
analyst to select the tuples of trades in $C$ to emphasize those that are
more relevant in a given application. The price to pay is double. First, the
maximum-score estimator maximizes a discontinuous function and converges slowly\footnote{%
The maximum-score estimator converges at a cubic-root rate.}. Second, the
underlying monotonicity property only holds for distributions of unobserved
heterogeneity that exclude nested logit models and random coefficients for
instance.

\section{Computation}\label{ch:structuraltumodels:sec:computation}
We now turn to the efficient evaluation of the stable matching and the associated utilities for given values of the parameters.
 In all of this section,  we consider any
distributional parameters $\bm{\beta}$ as fixed and we omit them from the notation.

\subsection{Solving for equilibrium with coordinate descent}\label{ch:structuraltumodels:sub:ipfp}
 First consider the determination of the  equilibrium matching patterns for a given matrix $\bm{\Phi}$.  In several important models, this can be done by adapting  formula~\eqref{ch:structuraltumodels:eq:mm:Estar}. A slight modification of the arguments that lead to this formula shows that for given $\bm{\Phi}$, maximizing the  following function yields the equilibrium utilities of all \idf{types}{Types}:
\begin{equation*}
\bar{F}(\bm{u},\bm{v}):= \sum_{x,y} \hat{\mu}_{xy}\Phi_{xy}
    -\dotprod{\hat{\bm{n}}}{\bm{u}}
    -\dotprod{\hat{\bm{m}}}{\bm{v}}-E^\ast\left(\bm{\Phi} -\bm{u}-\bm{v},-\bm{u},-\bm{v}\right).
\end{equation*}
\idf{Coordinate descent}{Coordinate descent} consists of maximizing $\bar{F}$ iteratively with respect to
the two argument vectors: with respect to $\bm{u}$ keeping $\bm{v}$ fixed, then with
respect to $\bm{v}$ keeping $\bm{u}$ fixed at its new value, etc.

Let $\bm{v}^{(t)}$ be the current value of $\bm{v}$. Minimizing $\bar{F}$ with respect to $\bm{u}$
for $\bm{v}=\bm{v}^{(t)}$ yields a set of $\left\vert \mathcal{X}\right\vert$
equations in $\left\vert \mathcal{X}\right\vert$ unknowns: $u_{x}^{(t+1)}$
is the value of $u_x$ that solves
\begin{align*}
\hat{n}_{x} &=\sum_{y\in \mathcal{Y}}\frac{\partial E^{\ast }}{\partial
z_{xy}}\left(\bm{\Phi}-\bm{u}-\bm{v}^{(t)},-\bm{u}, -\bm{v}^{(t)}\right) \\
&+\frac{\partial E^{\ast}}{\partial z_{x0}}\left(\bm{\Phi}-\bm{u}-\bm{v}^{(t)},-\bm{u}, -\bm{v}^{(t)}\right).
\end{align*}
These equations can in turn be solved coordinate by coordinate: we start
with $x=1$ and solve the $x=1$ equation for $u^{(t+1)}_1$ fixing $%
(u_2,\ldots,u_{\abs{\mathcal{X}}})=(u_2^{(t)},\ldots,u_{\abs{\mathcal{X}}%
}^{(t)})$; then we solve the $x=2$ equation for $u^{(t+1)}_2$ fixing $%
(u_1,u_3,\ldots,u_{\abs{\mathcal{X}}})=(u_1^{(t+1)},u_3^{(t)},\ldots,u_{%
\abs{\mathcal{X}}}^{(t)})$, etc. The convexity of the function $E^\ast$
implies that the right-hand side of each equation is strictly decreasing in
its scalar unknown, which makes it easy to solve.

The \idf{logit}{Logit} model constitutes an important special case in which
these equations can be solved with elementary calculations, for any joint
surplus matrix $\bm{\Phi}$. Define $S_{xy}:=\exp (\Phi_{xy}/2); a_{x}:=\exp
\left( -u_{x}\right)$; and $b_{y}:=\exp \left( -v_{y}\right)$. It is easy to
see that the system of equations that determines $\bm{u}^{(t+1)}$ becomes
\begin{equation*}
a_{x}^2 +a_{x}\sum_{y\in \mathcal{Y}}b^{(t)}_{y}S_{xy} =n_{x} \; \; \forall
x\in \mathcal{X}.
\end{equation*}
These are $\abs{\mathcal{X}}$ functionally independent quadratic equations,
which can be solved in closed-form and in parallel. Once this is done, a
similar system of independent quadratic equations gives $\bm{b}^{(t+1)}$ from $%
\bm{a}^{(t+1)}$. Note that $%
a^{(0)}_x=\sqrt{\hat{\mu}_{x0}}$ and $b^{(0)}_y=\sqrt{\hat{\mu}_{0y}}$ are
obvious good choices for initial values.

This procedure generalizes the
\idf{Iterative Proportional Fitting Procedure (IPFP)}{Iterative Proportional Fitting Procedure (IPFP)}, also known as \idf{Sinkhorn's
algorithm}{Sinkhorn's
algorithm}. It converges globally and very fast. Once the solutions $\bm{a}$ and $\bm{b}$ are obtained, the equilibrium
matching patterns for this $\bm{\Phi}$ are given by $\mu _{x0}=a_{x}^{2}$, $\mu
_{0y}=b_{y}^{2}$ and $\mu _{xy}=a_{x}b_{y}S_{xy}$.

\subsection{Gradient descent}
Suppose that the analyst has chosen to use~\eqref{ch:structuraltumodels:eq:mm:Estar} for estimation.
The simplest approach to maximizing the objective function is through \idf{gradient descent}{Gradient descent}.
Denoting $\bm{\alpha}=(\bm{\lambda},\bm{u},\bm{v})$, we start from a reasonable\footnote{In the  logit model, $u^{(0)}_x=-\log (\hat{\mu}_{x0}/\hat{n}_x)$
and $v^{(0)}_y=-\log (\hat{\mu}_{0y}/\hat{m}_y)$ are excellent choices of
initial values.} $\bm{\alpha}^{(0)}$
and we iterate:
\begin{equation*}
    \bm{\alpha}^{(t+1)}=\bm{\alpha}^{(t)}-\epsilon ^{(t)}\nabla F\left(\bm{\alpha}^{(t)}\right)
\end{equation*}%
where $\epsilon ^{(t)}>0$ is a small enough parameter. This gives%
\begin{align*}
u_{x}^{(t+1)}&=u_{x}^{(t)}+\epsilon ^{(t)}\left(n_{x}-N_x(\bm{\mu}^{(t)})\right)
\\
v_{y}^{(t+1)}&=v_{y}^{(t)}+\epsilon ^{(t)}\left(m_y -M_y(\bm{\mu}^{(t)})\right) \\
\lambda _{k}^{(t+1)}&=\lambda _{k}^{(t)}+\epsilon ^{(t)}\sum_{\substack{ %
x\in \mathcal{X}  \\ y\in \mathcal{Y}}}\left( \mu _{xy}^{(t)}-\hat{\mu}%
_{xy}\right) \phi _{xy}^{k},
\end{align*}%
denoting $\bm{\mu}^{(t)}$ the result of plugging $(\bm{u}^{(t)},\bm{v}^{(t)},\bm{\lambda}^{(t)})$
into~(\ref{ch:structuraltumodels:def_pi_theta}).

This algorithm has a simple intuition: we adjust $u_{x}$ in proportion of
the excess of $x$ \idf{types}{Types}, $v_{y}$ in proportion of the excess of $y$ types,
and $\bm{\lambda}$ in proportion of the mismatch between the $k$-th moment
predicted by $\bm{\alpha}$ and the observed $k$-th moment.

\subsection{Hybrid Algorithms}
The approaches in the previous two subsections can also be combined. \cite{carlier2020sista} suggest alternating between
coordinate descent steps on $\bm{u}$ and $\bm{v}$ and gradient descent steps on $%
\bm{\lambda}$. In the \idf{logit}{Logit}
model, this would combine the updates
\begin{equation*}
\left\{
\begin{array}{l}
\left(a^{(t+1)}_{x}\right)^2 +a^{(t+1)}_{x}\sum_{y\in \mathcal{Y}%
}b^{(t)}_{y}S^{(t)}_{xy} =n_{x} \\
\left(b^{(t+1)}_{y}\right)^2 +b^{(t+1)}_{y}\sum_{x\in \mathcal{X}%
}a^{(t+1)}_{y} S^{(t)}_{xy} =m_y \\
\lambda _{k}^{(t+1)}=\lambda _{k}^{(t)}+\epsilon ^{(t)}\sum_{\substack{ x\in
\mathcal{X}  \\ y\in \mathcal{Y}}}\left(a_{x}^{(t+1)}b_{y}^{(t+1)}
S_{xy}^{(t)} -\hat{\mu}_{xy}\right) \phi_{xy}^{k}%
\end{array}%
\right.
\end{equation*}
where $S_{xy}^{(t)} =\exp(\sum_{k=1}^K \phi_{xy}^k \lambda^{(t)}_k/2)$.

A proof of convergence of hybrid algorithms is given in~\cite%
{carlier2020sista}, in a more general setting that allows for model
selection based on penalty functions.

\section{Other Implementation Issues}

Let us now very briefly discuss three issues that often crop up in
applications.

\subsection{Continuous Types}

While we modeled types as discrete-valued in this chapter, there are
applications where this is not appropriate. It is possible to
incorporate continuous types in a  separable model that feels very similar to the  logit model of Section~\ref{ch:structuraltumodels:sub:mlogit}. The idea is to model the choice of possible partners as generated by
the points of a specific \idf{Poisson process}{Poisson process}.  An interesting special case has a bilinear \idf{joint surplus function}{Joint surplus function} $\Phi(x,y) = x^{\top} A y$.  It is easy to see that at the optimum, the Hessian of the logarithm of the matching patterns equals $A$ everywhere:  for all $x\in \mathbb{R}^{d_{x}}$
and  $y\in \mathbb{R}^{d_{y}}$,
\begin{equation*}
\frac{\partial ^{2}\ln \mu}{\partial x \partial y} \left( x,y\right)
=\frac{A}{2}.
\end{equation*}
As a consequence, the  model is overidentified and therefore testable. Among other things, it makes it possible to test for the rank of the matrix $A$. If it is some $r<\min(d_x, d_y)$, then one can identify the ``salient'' combination of types that generate the joint surplus.

If moreover the distribution $P$ of $x$ and the distribution $Q$ of $y$  are Gaussians,  that the \idf{optimal matching}{Optimal matching} $\left(X,Y\right)$ is a Gaussian vector whose
distribution can be obtained in closed form. Suppose for instance that $%
d_{x}=d_{y}=1$; $P=\mathcal{N}\left( 0,\sigma _{x}^{2}\right)$; $Q=\mathcal{N%
}\left( 0,\sigma _{y}^{2}\right)$; and $\Phi \left( x,y\right) =axy$, Then
at the optimum $V X =\sigma _{x}^{2}$, $VY =\sigma _{y}^{2}$, and $%
\mbox{corr}\left( X,Y\right) =\rho$ where $\rho$ is related to $a$ by%
\begin{equation*}
a\sigma _{x}\sigma _{y}=\frac{\rho }{1-\rho ^{2}}.
\end{equation*}

\subsection{Using Several Markets}\label{ch:structuraltumodels:sub:Using Several Markets}
We have focused on the case when the
analyst has data on one market. If data on several markets is available;
matches do not cross market boundaries; and some of the primitives of the
model coincide across markets, then this can be used to relax the conditions
necessary for \idf{identification}{Identification}.

As an example, \cite{csw:17} pooled Census data on thirty cohorts in the US
in order to study the changes in the marriage returns to education. To do
this, they assumed that the supermodularity module of the function $\Phi$
changed at a constant rate over the period.

\cite{foxyanghsu} show how given enough markets, one can identify the
distribution of the \idf{unobserved heterogeneity}{Unobserved heterogeneity} if it is constant across
markets.

\subsection{Using Additional Data}

In applications to the labor market for instance, the analyst often has some
information on transfers---wages in this case. This information can be used
in estimating the underlying matching model. It is especially useful if it
is available at the level of each individual match. Aggregate data on
transfers has more limited value \citep{salanieidentobstransfers:15}.

\section{Notes}
Matching with perfectly transferable utility was introduced by~\cite{kb:57} and its theoretical properties were elucidated by  \cite{ss:72}. \cite{becker:73, becker:74} made it the cornerstone of his analysis of marriage.
Sections~\ref{ch:structuraltumodels:sec:ident} and~\ref{ch:structuraltumodels:sec:estimation} of this chapter are  based on
the approach developed in~\cite{cupid:20}. The extension of the  logit model to continuous types was proposed by~\cite{dg:14}, following~\cite%
{dagsvik:00}.  They applied it to study how the joint surplus from marriage depends on the Big Five psychological traits of the partners. \cite{grst:20} combine continuous and discrete types to model mergers between European firms. The results for the bilinear Gaussian models appear in~\cite{bojilov:16}.

 The maximum-score method for matching models   was proposed by~\cite%
{fox:identmatchinggames}, taking inspiration from~\cite{ManskiMaxScore:75}'s classic paper on one-sided discrete choice models. \cite%
{bajarifox} used this estimator to study the FCC spectrum auctions. \cite{graham:hdbk,graham:hdbkerrata} proved Theorem~\ref{ch:structuraltumodels:thm:maxscore:exch} for independent and identically distributed variables and~\cite{fox:qe18} extended it to exchangeable variables.

\bibliographystyle{cambridgeauthordate}
\bibliography{Matching}

\section*{Appendix A: reminders on convex analysis}\label{ch:structuraltumodels:convexanalysis}

 We focus here on  the  results that our chapter relies on. For an economic interpretation in terms of
matching, see Chapter~6 of~\cite{otme}.

\bigskip

In what follows, we consider a convex function $\varphi :\mathbb{R}%
^{n}\rightarrow \mathbb{R\cup }\left\{ +\infty \right\} $ which is not
identically $+\infty $. If $\varphi $ is differentiable at $x$, we denote
its \emph{gradient} at $x$ as the vector of partial derivatives, that is $%
\nabla \varphi \left( x\right) =\left( \partial \varphi \left( x\right)
/\partial x_{1},\ldots ,\partial \varphi \left( x\right) /\partial
x_{n}\right) $. In that case, one has for all $x$ and $\tilde{x}$ in $%
\mathbb{R}^{n}$
\begin{equation*}
\varphi \left( \tilde{x}\right) \geq \varphi \left( x\right) +\nabla \varphi
\left( x\right) ^{\top }\left( \tilde{x}-x\right) .
\end{equation*}

Note that if $\nabla \varphi \left( x\right) $ exists, then it is the only
vector $y\in \mathbb{R}^{n}$ such that
\begin{equation}
\varphi \left( \tilde{x}\right) \geq \varphi \left( x\right) +y^{\top
}\left( \tilde{x}-x\right) \;\;\forall \tilde{x}\in \mathbb{R}^{n},
\label{ch:structuraltumodels:subdiff-rel}
\end{equation}%
indeed, setting $\tilde{x}=x+te_{i}$ where $e_{i}$ is the $i$th vector of
the canonical basis of $\mathbb{R}^{n}$, and letting $t\rightarrow 0^{+}$
yields $y_{i}\leq \partial \varphi \left( x\right) /\partial x_{i}$, while
letting $t\rightarrow 0^{-}$ yields $y_{i}\geq \partial \varphi \left(
x\right) /\partial x_{i}$. This motivates the definition of the \emph{%
subdifferential} $\partial \varphi \left( x\right) $ of $\varphi $ at $x$ as
the set of vectors $y\in \mathbb{R}^{n}$ such that relation~(\ref{ch:structuraltumodels:subdiff-rel}) holds. Equivalently, $y\in \partial \varphi \left( x\right) $
holds if and only if
\begin{equation*}
y^{\top }x-\varphi \left( x\right) \geq \max_{\tilde{x}}\left\{ y^{\top }%
\tilde{x}-\varphi \left( \tilde{x}\right) \right\}
\end{equation*}%
that is, if and only if
\begin{equation*}
y^{\top }x-\varphi \left( x\right) =\max_{\tilde{x}}\left\{ y^{\top }\tilde{x%
}-\varphi \left( \tilde{x}\right) \right\} .
\end{equation*}

The above development highlights a special role for the function $\varphi
^{\ast }$ appearing in the expression above%
\begin{equation*}
\varphi ^{\ast }\left( y\right) =\max_{\tilde{x}}\left\{ y^{\top }\tilde{x}%
-\varphi \left( \tilde{x}\right) \right\}
\end{equation*}%
which is called the \emph{Legendre-Fenchel transform} of $\varphi$. By
construction,
\begin{equation*}
\varphi \left( x\right) +\varphi ^{\ast }\left( y\right) \geq y^{\top }x.
\end{equation*}%
This is called Fenchel's inequality; as we just saw, it is an equality if
and only if $y\in \partial \varphi \left( x\right) $. In fact, the
subdifferential can also be defined as
\begin{equation*}
\partial \varphi \left( x\right) =\arg \max_{y}\left\{ y^{\top }x-\varphi
^{\ast }\left( y\right) \right\} .
\end{equation*}

Finally, the double Legendre-Fenchel transform of a convex function $\varphi$
(the transform of the transform) is simply $\varphi$ itself. As a
consequence, the subgradients of $\varphi$ and $\varphi^\ast$ are inverses
of each other. In particular, if $\varphi$ and $\varphi^\ast$ are both
differentiable then
\begin{equation*}
(\nabla \varphi)^{-1}=\nabla \varphi ^{\ast}.
\end{equation*}
To see this, remember that $y\in \partial \varphi \left( x\right) $ if and
only if $\varphi \left( x\right) +\varphi ^{\ast }\left( y\right) =y^{\top
}x $; but since $\varphi^{\ast\ast}=\varphi$, this is equivalent to $\varphi
^{\ast \ast }\left( x\right) +\varphi ^{\ast }\left( y\right) =y^{\top }x$,
and hence to $x\in \partial \varphi ^{\ast }\left( y\right)$. As a result,
the following statements are equivalent:

\begin{enumerate}
\item[(i)] $\varphi \left( x\right) +\varphi ^{\ast }\left( y\right)
=x^{\top }y$;

\item[(ii)] $y\in \partial \varphi \left( x\right)$;

\item[(iii)] $x\in \partial \varphi ^{\ast }\left( y\right) $.
\end{enumerate}

\section*{Appendix B: asymptotic distribution of the logit moment-matching estimator}

In this appendix, we provide the explicit formulas for the asymptotic distribution of the estimator of the matching surplus
in the logit model of Section~\ref{ch:structuraltumodels:sub:logitEstim}. The asymptotic distribution of the estimator $\hat{\bm{\alpha}}$ is easy to derive by totally
differentiating the first order conditions $F_{\bm{\alpha}}(\hat{\bm{\alpha}},\hat{\bm{\pi}%
})=0$. This yields
\begin{equation*}
\bm{\alpha} \sim \mathcal{N}\left( 0,\frac{V_{\bm{\alpha}}}{N_{h}}\right)
\end{equation*}%
where%
\begin{equation*}
V_{\bm{\alpha}}=\left( F_{\bm{\alpha}\bm{\alpha}}\right) ^{-1}F_{\bm{\alpha}\bm{\pi} }V_{\pi
}F_{\bm{\alpha}\bm{\pi} }^{\top }\left( F_{\bm{\alpha}\bm{\alpha}}\right) ^{-1}.
\end{equation*}%
In this formula, $V_{\pi}$ is as in~\eqref{ch:structuraltumodels:eq:Vpi}
and the $F_{ab}$ represent the blocks of the Hessian of $F$ at $(\hat{\bm{\alpha}}%
,\hat{\bm{\pi}})$. Easy calculations show that $F_{\bm{\alpha}\bm{\alpha}}$ in turn
decomposes into%
\begin{equation*}
\begin{pmatrix}
F_{\bm{uu}} & F_{\bm{uv}}=\left( \frac{\pi _{xy}^{\lambda }}{2}\right) _{xy} &
F_{\bm{u\lambda}}=-\frac{1}{2}\left( \sum_{y}\pi _{xy}^{\lambda }\phi
_{xy}^{k}\right) _{xk} \\
. & F_{\bm{vv}} & F_{\bm{v\lambda}}=-\frac{1}{2}\left( \sum_{x}\pi _{xy}^{\lambda
}\phi _{xy}^{k}\right) _{yk} \\
. & . & F_{\bm{\lambda\lambda}}=\frac{1}{2}\left( \sum_{x,y}^{\lambda }\hat{\pi}%
_{xy}\phi _{xy}^{k}\phi _{xy}^{l}\right) _{kl}%
\end{pmatrix}%
\end{equation*}%
where
\begin{equation*}
F_{\bm{uu}}=\mbox{diag}\left( \left( \frac{1}{2}\sum_{y}\pi _{xy}^{\lambda }+\pi
_{x0}^{\lambda }\right) _{x}\right) \text{ and }F_{\bm{vv}}=\mbox{diag}\left(
\left( \frac{1}{2}\sum_{x}\pi _{xy}^{\lambda }+\pi _{0y}^{\lambda }\right)
_{y}\right).
\end{equation*}
Moreover,
\begin{equation*}
F_{\bm{\theta} \bm{\pi} }=%
\begin{pmatrix}
\left( 1_{\mathcal{Y}}^{\top }\otimes I_{\mathcal{X}}\right) & I_{X} & 0 \\
\left( I_{\mathcal{Y}}\otimes 1_{\mathcal{X}}^{\top }\right) & 0 & I_{Y} \\
\left( -\phi _{xy}^{k}\right) _{k,xy} & 0 & 0%
\end{pmatrix}%
.
\end{equation*}
 Once the estimates $\hat{\bm{\alpha}}$ are obtained, we can apply~\eqref{ch:structuraltumodels:def_pi_theta} to compute the estimated matching patterns:
\begin{equation*}
    \left\{
    \begin{array}{c}
    \mu^{\hat{\alpha}} _{xy}=\exp\left(\Phi^{\hat{\lambda}}_{xy}-\hat{u}_x-\hat{v}_y)/2\right) \\[3mm]
    \mu^{\hat{\alpha}} _{x0}=\exp\left(-\hat{u}_x\right)\\[3mm]
    \mu^{\bm{\alpha}} _{0y}=\exp\left(-\hat{v}_y\right).
    \end{array}%
    \right.
    \end{equation*}

\end{document}